\documentclass[galaxies,article,accept,moreauthors,pdftex]{Definitions/mdpi}
\firstpage{1}
\pubvolume{7}
\issuenum{4}
\articlenumber{84}
\pubyear{2019}
\copyrightyear{2019}
\history{Received: 19 July 2019; Accepted: 11 September 2019; Published: date}
\updates{yes} 
\setitemize{parsep=6pt,itemsep=0pt,leftmargin=*,labelsep=5.5mm}
\setenumerate{parsep=6pt,itemsep=0pt,leftmargin=*,labelsep=5.5mm}
\setlist[description]{itemsep=0mm} 

\Title{Planets in Binaries: Formation and Dynamical~Evolution}

\Author{Francesco Marzari $^{1,}$* and Philippe Thebault $^{2}$}

\AuthorNames{Francesco Marzari, Philippe Thebault}

\address{%
$^{1}$ \quad Department of Physics and Astronomy, University of Padova, 35122 Padova PD, Italy\\
$^{2}$ \quad Observatoire del Paris Meudon, 75014  Paris, France; Philippe.Thebault@obspm.fr}

\corres{Correspondence: marzari@pd.infn.it }

\abstract{Binary systems are very common among field stars, yet the vast majority of known exoplanets have been detected around single stars.
While this relatively small number of planets in binaries is probably partly due to strong observational biases, there is, however, statistical evidence that planets are indeed less frequent in binaries with separations smaller than 100 au, strongly suggesting that the presence of a close-in companion star has an adverse effect on planet formation.
It is indeed possible for the gravitational pull of the second star to
affect all the different stages of planet formation, from proto-planetary disk formation to dust accumulation into 
planetesimals, to the accretion of these planetesimals into large planetary embryos and, eventually, the final 
growth of these embryos into planets. For the crucial planetesimal-accretion phase, the complex coupling between dynamical perturbations
from the binary and friction due to gas in the proto-planetary disk suggests that planetesimal accretion
might be hampered due to increased, accretion-hostile impact velocities.
Likewise, the interplay between the binary's secular perturbations and mean motion resonances lead to unstable
regions, where not only planet formation is inhibited, but where a 
massive body would be ejected from the system on a hyperbolic orbit. The 
amplitude of these two main effects is different for S- and P-type 
planets, so that a comparison between the two populations might 
outline the influence of the companion star on the planet formation process.
Unfortunately, at present the two populations (circumstellar or circumbinary) 
are not known equally well and different biases and uncertainties prevent 
a quantitative comparison. 
We also highlight the long-term dynamical evolution of both S and P-type systems and focus on how these different evolutions influence the final architecture of planetary systems in binaries.
}


\begin{document}

\section{Introduction}

\textls[-10]{In a binary star system, the~gravity of the companion star may 
strongly influence both the formation of planets and their subsequent
dynamical evolution. The~final architecture of a system  strongly depends
on how the efficiency of the accretion process 
is affected by 
the presence of a close massive body and on how 
the subsequent evolution 
due to migration by interaction with the circumstellar disk, 
tidal interaction with the star, planet--planet scattering 
or Kozai mechanism, is~altered by the second star. 
The differences in planet growth and dynamics in binaries, with~respect
to single stars, should show up
in the fraction of binary systems 
harboring planets and in their present dynamical 
configurations. However, to~interpret properly what the 
data tell us, we need to develop models that can explain 
the specificities of these planetary systems compared to those
around single~stars.}

We will focus here on the main characteristics of planets in both 
S and P--type orbital configurations which significantly 
differ in terms of perturbations by the companion star. 
In the S-type configuration, where planets orbit one of the star of the system, the~most critical regions for 
planet formation are the outer ones where the gravity of the 
second star may excite large eccentricities, potentially leading  
to instability via resonance superposition and, in~case 
of orbital misalignment, may also be involved in fast Kozai 
cycles with large eccentricity/inclination variations. 
The inner 
regions of the circumstellar disk appears then most promising
for a 'regular' core--accretion growth of planets and for subsequent
stable orbital behavior. 
In the 
case of P--type configurations, where the planets orbit around both stars, 
perturbations are the strongest in the inner regions close to the central binary.
In this configuration, instability is expected close 
to the center of the system, while planet formation should 
safely occur in the outer regions of 
the circumbinary disk. In~this case, planet migration may play
a major role in bringing outer planets close to the 
binary. 

We expect that these diversities in the dynamics of S and P--type
trajectories lead to distinct populations of planets, differing
both from a physical and dynamical point of view. The~diversity would
be a consequence of the different effect the companion star has
on the formation of planets and on their subsequent dynamical evolution. 
Unfortunately, modeling 
the formation and evolution of planets in binaries is 
a very complex task due to the wide parameter
space to explore, which 
include the binary mass ratio and the semi-major
axis, eccentricity and inclination of the binary orbit. 
A change in these parameters may increase/decrease the 
dynamical perturbation at different locations in the 
planet forming disk affecting all the stages of the 
evolution of the system, from~proto-planetary disk formation 
to the final dynamical evolution of the system. Due to this
variety of binary configurations it is an impossible task 
to outline a 'standard model' of planet formation in binaries 
as opposed to that around single stars. In~some configurations 
of the binary the growth of planets may follow a path similar
to that around single stars, while in others, due to the 
companion star stronger influence, the~evolution may be completely 
different, either favoring the formation of more/less massive
planets or inhibiting the presence of stable planets at 
given radial distances from the primary star or from the 
baricenter of the pair, or~leading to planets which are more 
eccentric or more inclined with respect to those around single stars. 
In addition, the~two populations of planets, either circumstellar 
or circumbinary, are not equally well known since more than 120 planets
on S-type orbits are known, as~compared to only a dozen of P-type planets. In~addition, the~majority of confirmed S-type planets were detected by radial velocities, whereas circumbinary planets were mostly
 detected by~transits.

In this paper we summarize the properties and peculiarities of 
planets in binaries and relate them to the formation process 
and the subsequent complex dynamics caused by the companion~star. 

\section{Planets in S--Type~Orbits}
\unskip

\subsection{Observational~Constraints}
\unskip

\subsubsection{Adverse~Biases}

As of July 2019, 122 exoplanet host stars, out of 3055, are known to be part of binary or multiple systems. 
In an ideal world, the~effect of binarity on exoplanet presence could be directly inferred from this raw data, by~simply estimating the binarity rate among exoplanet hosts, as~well as the distribution of separations for these planet-bearing binaries, and~compare it to the corresponding statistics for field stars. Unfortunately, such a simple comparison is almost certainly~misleading. 

\begin{figure}[H]
\centering
  \includegraphics[scale=0.55]{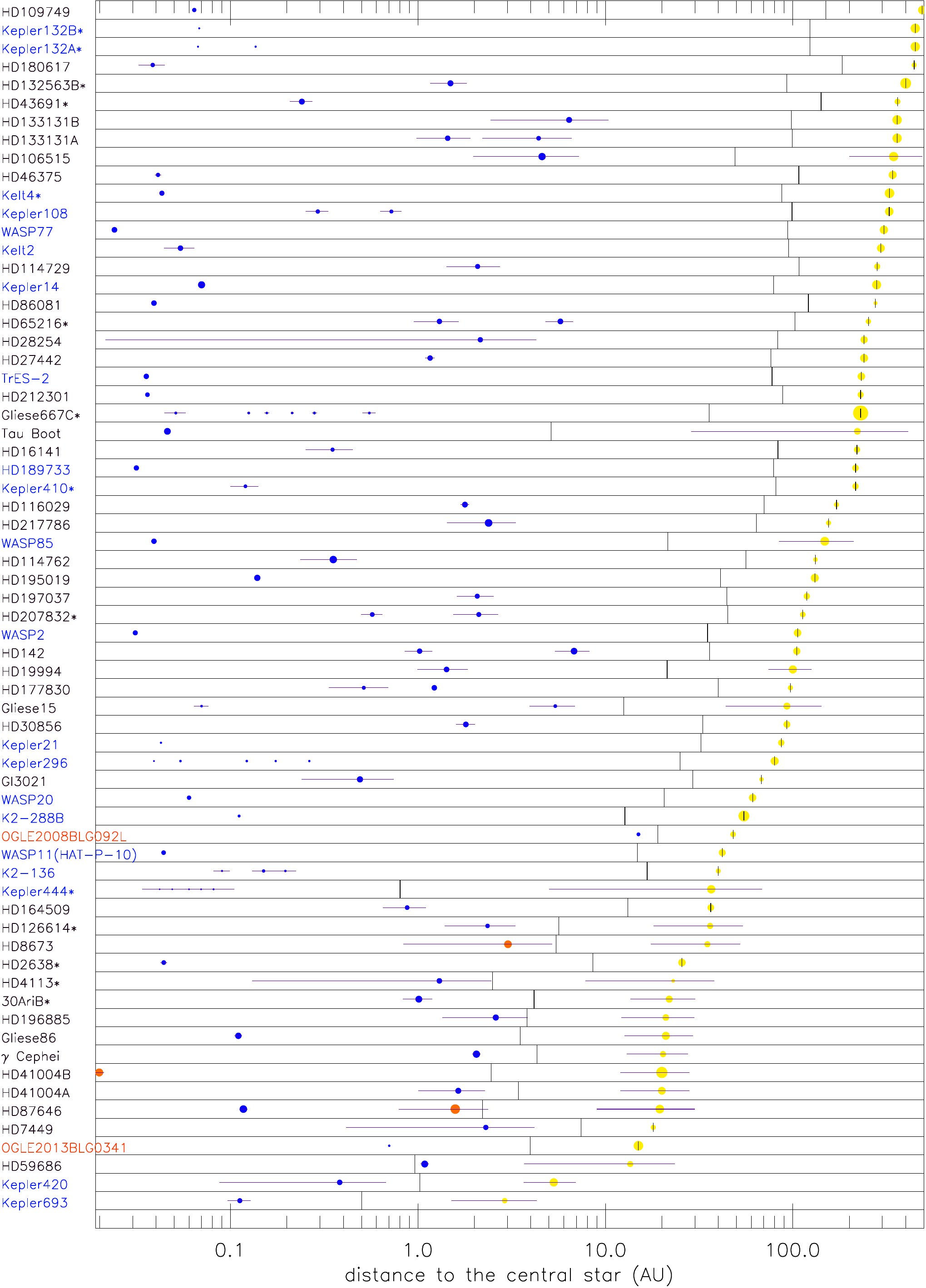}
  \caption{All circumstellar (S-type orbits, as~opposed to circumbinary P-type orbits) planet-hosting binaries with separation $\leq$500\,au (as of July 2019). Companion stars are displayed as yellow circles, whose radius is proportional to $(M_2/M_1)^{1/3}$. Planets are marked as blue circles whose radius is proportional to $(m_{pl})^{1/3}$. The~horizontal lines represent the radial excursion of the planet and star orbits (when they are known). When the binary orbit is known, the~displayed distance is the semi-major axis, otherwise it is the projected current separation. The~short vertical lines correspond to the outer limit of the orbital stability region around the primary  estimated by \mbox{\citet{holman99}} assuming prograde orbits and coplanarity (for the seemingly ``unstable''  planet HD59686Ab, a~retrograde orbit was suggested as a possible stable solution \citep{trifo2018}. Planets detected by the radial velocity method are written in black, planets detected by transit are in blue, and~planets detected by other methods are in~red. } 
 \label{bindistrib}
\end{figure}

Taken at face value, 122 out of 3055 exoplanet-host stars would indeed yield a very low binary fraction of $\sim$4\%, as~compared to approximately 50\% for solar type field stars \citep{duma,ragha10}, which would imply that binarity has a very strong adverse effect on the presence of exoplanets. However, this fraction should be taken with great caution because it suffers from very strong biases. The~first issue is that radial velocity (RV) surveys, which account for roughly half of the discovered S-type exoplanets, are~in general strongly biased against binaries, and~often discard any binaries with separation $\leq$2$''$ that can fall to within the spectrograph slit \citep{eggen10,Ngo17}. An~additional problem is that these biases are never well characterized and might differ from one survey to the other. One of the few attempts at investigating the effect of binarity on exoplanet occurrence within one coherent RV survey is that of Eggenberger and her team \citep{eggen04,eggen07,eggen10}, which compared the binarity fraction of a sample of exoplanet host stars to a control sample of stars with no RV planets, and~found these fraction to be $\sim$5.5\% and $\sim$13.7\%, respectively~\citep{eggen10}, thus confirming the global adverse effect of binarity on exoplanet presence. Note, however, that, despite these adverse selection effects, the~vast majority of exoplanets in tight binaries (of separation $\leq$50\,au) were detected by radial velocity surveys (Figure \ref{bindistrib}).

Transit surveys, which account for $\sim$40\% of S-type exoplanets, do not suffer from these adverse selection effects. However, the~stars they are targeting, in~particular those of the NASA Kepler telescope, are generally more distant and have often not been vetted for binarity yet, so that a significant fraction might be wrongly labelled as singles. To~alleviate this problem, several adaptive optic (AO) surveys were carried out in recent years, which detected a large number of companion stars to known transit-planet hosts \citep{wang14a,wang14b,ziegler17,Ngo17} \footnote{Such AO searches for stellar companions have also been performed for RV-planet hosts, and~have increased the binary fraction among them, but~not by a factor as large as for transit exoplanet hosts \citep{eggen07,mugrauer15,ginski12,ginski16}.}. Note, however, that most of these surveys were targeting non-confirmed Kepler Objects of Interest (KOIs), so that the actual binarity fraction among \emph{confirmed} exoplanets stays relatively low. This is why, even if transit detections represent more than 70\% of known exoplanets, they only account for $\sim$40\% of confirmed exoplanets in~binaries. 

However, despite their respective shortcomings and limitations, 
RV and transit detections have uncovered some general and relatively robust trends regarding exoplanets in~binaries.

\subsubsection{Characteristics of Exoplanets in~Binaries}

An investigation that is in principle not sensitive to target selection biases is the comparison between the exoplanet populations in single and multiple~systems. 

A trend that was identified early on, albeit on a relatively limited sample of RV-detected systems, is~the high fraction of massive close-in planets (``hot Jupiters'') in tight ($\rho\leq100\,$au) binaries as compared to their fraction in wide binaries or single stars \citep{zuck02,eggen04,desidera07}. More recent studies, considering larger samples of exoplanet systems, confirmed that there is indeed a clear difference regarding the populations of hot Jupiters in multiple and single stars, but~the picture that emerges is different from the one envisaged early on. 
On the one hand, contrary to early estimates, the~tendency towards more frequent hot-Jupiters in tight binaries is no longer found. Indeed,~reference \cite{Ngo16} find that, for~separations $\leq$50\,au, the~stellar multiplicity rate for hot-Jupiters is in fact 4 times lower than that of field stars, which~is approximately the same ratio as the one derived, for~the same separation range, for~the general population of exoplanet hosts \citep{Kraus16}. On~the other hand, there is a clear tendency towards a very high fraction of hot-Jupiter in moderate-to-wide binaries. For~the $50\leq\rho\leq2000\,$au separation range, reference~\cite{Ngo16} indeed find a multiplicity fraction that is 2.9 times \emph{higher} for hot-Jupiter hosts than for field stars, a~result which was confirmed by~\cite{fonta19} for close-in companions up to the brown dwarf~regime.

Regarding the population of exoplanets with semi-major axis $a_P\geq0.1\,$au, however, there does not seem to be any statistical difference between single and multiple stellar systems \citep{Ngo17}, neither regarding planet masses nor their orbital~parameters.

\subsubsection{Multiplicity Rate of Exoplanet~Hosts}

\begin{figure}[H]
\centering
  \includegraphics[width=0.8\columnwidth]{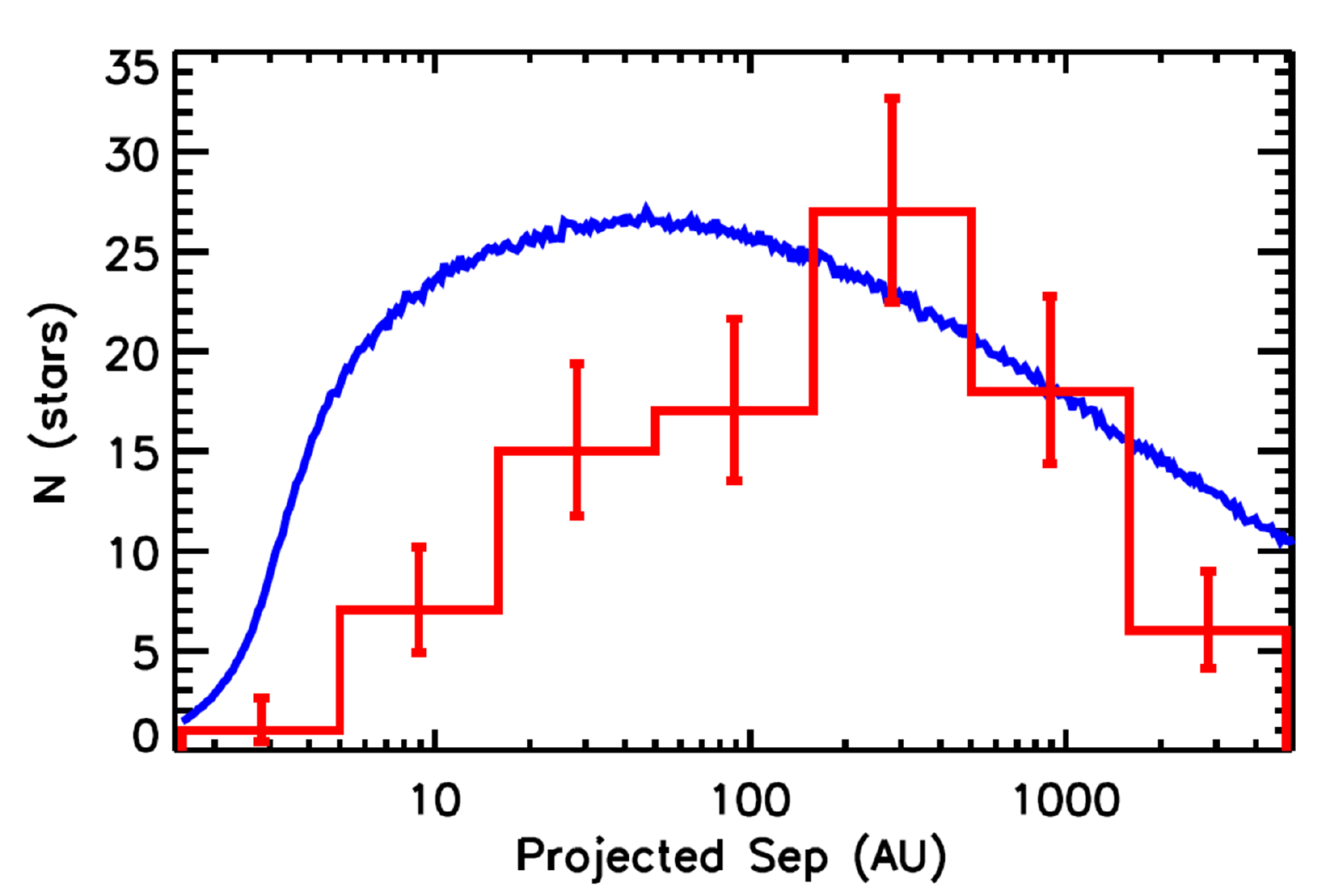}
  \caption{Red histogram: Marginalized distribution  $N(\rho)$ of projected separations for all stellar companions to S-type exoplanet-host stars found by~\cite{Kraus16} for a large sample of Kepler Objects of Interest (KOIs). The~blue line represents the predicted distribution if binaries were drawn from the distribution reported by~\cite{ragha10}. Taken from~\cite{Kraus16}, courtesy of the Astrophysical~Journal. }
 \label{Krausdistrib}
\end{figure}

One of the most obvious trends that shows up on the global raw distribution of separations $\rho$ of exoplanet-hosting binaries is a depletion of close-in companions ($\rho \leq100\,$au) as compared to the distribution of binary separations for solar-type field stars, which peaks around 30--50\,au (see~Figure~\ref{Krausdistrib}). This trend did already show up in the results of AO surveys looking for companions to RV-exoplanet hosts \citep{desidera07,eggen11}, but~the complexity of selection biases in RV targets prevented an undisputed confirmation as to its reality. It is the large-scale surveys looking for companions to transit exoplanet hosts that allowed to confirm this trend. The~Doppler and AO--based search of~\cite{wang14a} and the AO survey of~\cite{Kraus16} showed that, while the multiplicity fraction is statistically indistinguishable between planet and non-planet hosts for binary separations $\geq100\,$au, there is a very strong depletion of tighter companions for exoplanet hosts,
comprised between a factor $\sim$4 and $\sim$15.  Reference \cite{Kraus16} estimated that the distributions of close-in companions to exoplanet-hosting primary stars can be fitted by considering that the separation distribution for field stars is diminished by a factor $\sim$3 below a limiting separation $\rho_{cut}=47\,$au. In~the absence of selection biases, this is equivalent to saying that the planet occurrence rate is 3 times smaller in $\rho\leq47\,$au binaries than for single stars.
Note, however, that this issue is not fully settled yet. As~an example, while~\cite{wang14b} confirm a strong drop of planet occurrences in close-in binaries, they also found that this lower occurrence rate extends, albeit less pronounced, up~to separations of $\sim$1500\,au. On~the contrary, the~recent AO survey by~\cite{matson18} did not find any depletion of $\rho\leq50\,$au binaries in K2 exoplanet host stars. These conflicting results might be due to the fact that, for~Kepler targets, the~ $\rho\leq50\,$au separation domain is at the limit of what is feasible with AO companion searches.
Note also that these statistics about multiplicity rates in planet-hosts are only valid for the type of exoplanets that are detectable by given observing facilities.  As~an example, the~planet occurrence suppression rate for tight binaries found by~\cite{wang14b} or~\cite{Kraus16} apply to the short period (typically less than 100 days) planets investigated by Kepler, while the results regarding companions to RV-exoplanet hosts found by~\cite{desidera07,eggen11} are less biased in orbital-periods but strongly biased towards large mass~planets.

\begin{figure}[H]
\centering
  \includegraphics[width=0.8\columnwidth]{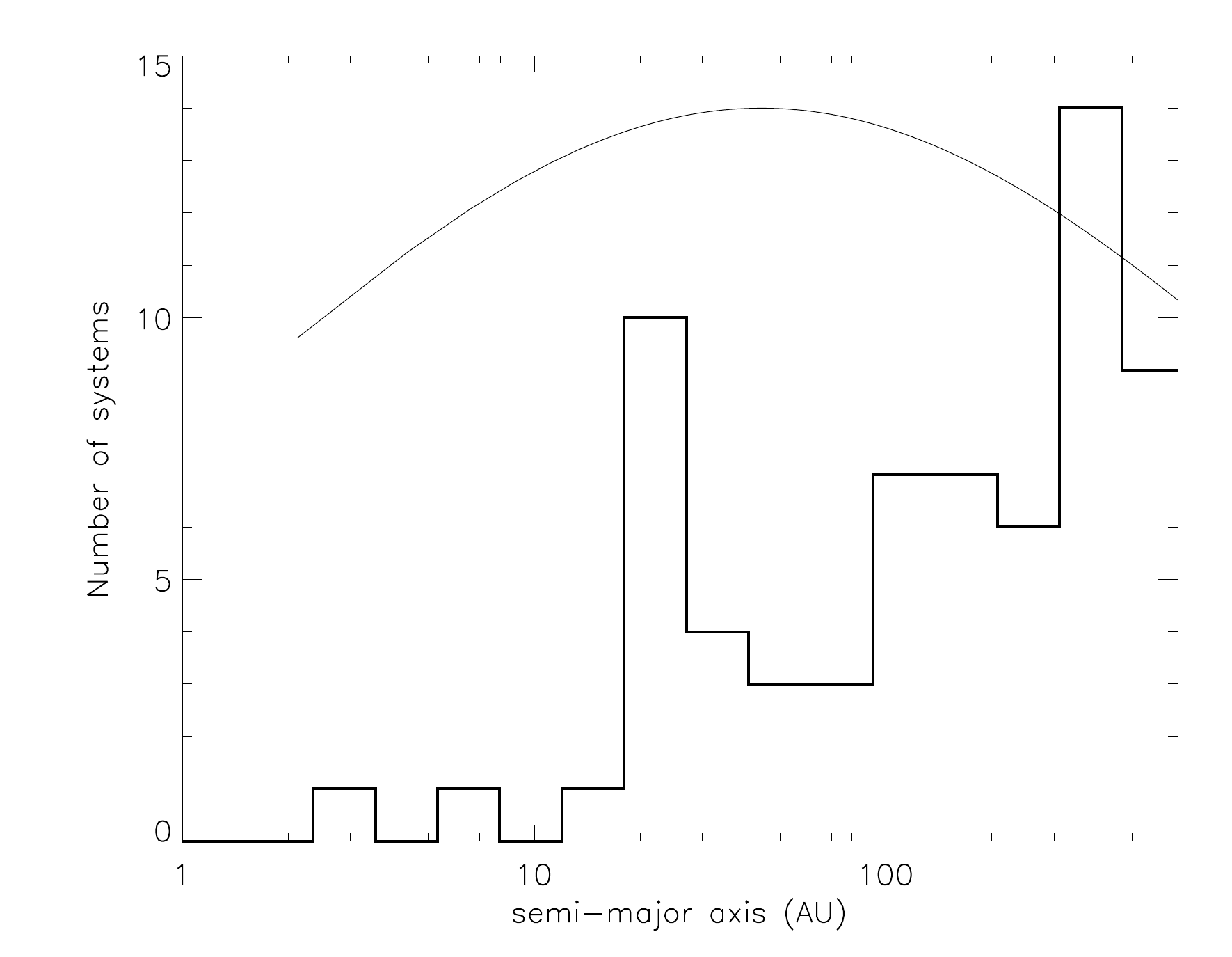}
  \caption{Distribution of the semi-major axis of all S-type exoplanet-hosting binaries with a projected separation up to 500 au. When the semi-major axis $a_b$ is not known and only the projected separation $\rho$ is available, $_ba$ is estimated through the statistical relation $log(a_b)=log(\rho)+0.13$ \citep{duma,ragha10}. The~solid-line curve represents the normalized $a_b$ distribution for field stars derived by~\cite{ragha10}.}
 \label{rawdistrib}
\end{figure}

There remains one potentially significant feature that shows up on the raw distribution of binary separations of exoplanet-hosts that has not been investigated yet, which is the apparent accumulation of exoplanet-host binaries around a $\rho \sim20\,$au separation (Figure \ref{rawdistrib}). Given the still limited number (9) of systems concerned it is still difficult to assess the reality of this feature, but~should it be confirmed then it would put very strong constraints on the formation and evolution of planets in multiple systems. At~any rate, the~very presence of planets in such ``extreme'' systems, where a stellar companion is located at the same position as Uranus in the solar system, is a challenge to the standard model of planet formation, especially when considering that some of these planets (HD196885b, $\gamma$ Cephei b, HD41004Ab) are both very massive and beyond 1.5\,au from their central~star.

\subsection{Planet Formation in~Binaries}

As shown in the previous section, there seems to be reliable observational evidence that a close-in stellar companion ($\leq$50\,au) has a detrimental effect on the presence of an exoplanet around a host star. In~addition, for~the few exoplanets that have been found to inhabit  ``extreme'' systems with binary separation of the order of 20\,au or less, it is very unlikely that their formation process could have proceeded without being strongly affected by the perturbations of the companion star. 
At~any rate, studying planet formation in binaries can be used as a test bench for planet formation models, by~confronting them to an unusual and sometimes ``extreme'' environment where some crucial parameters might be pushed to extreme~values.

For a detailed discussion on the issue of planet formation in binaries, we refer the reader to the review by~\cite{theb15}. We shall here highlight the main aspects of this very complex~issue.

\subsubsection{Early~Stages}

\begin{figure}[H]
\centering
  \includegraphics[width=0.8\columnwidth]{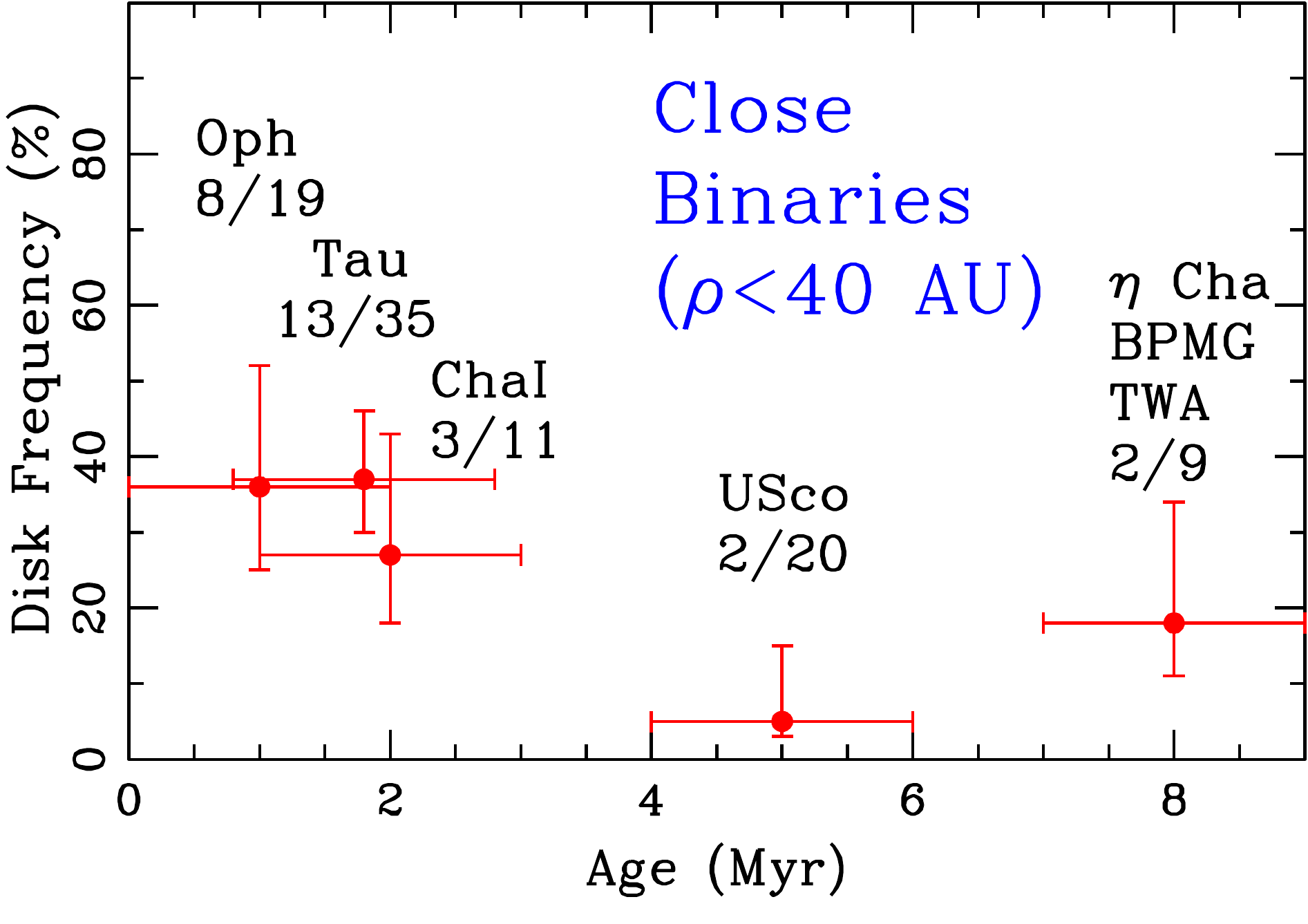}
  \caption{Frequency of disk--bearing tight binaries (with projected separation $\rho\leq40\,$au) as a function of age in several young stellar associations (from \cite{kraus12}, courtesy of the Astrophysical Journal).}
 \label{Krausdisc}
\end{figure}

The presence of a close-in stellar companion can have an effect early on in the planet formation, by~tidally truncating the circumstellar proto-planetary disk. This truncation radius is typically located at 1/3 to 1/4 of the binary's separation \citep{arty94,savo94}, which means that, if~we assume that the dense parts of a proto-planetary disk have a typical size of $\sim$50--100\,au, then this truncation effect should affect all binaries with separation $\leq$150--200\,au.   Observations show that disks are indeed less frequent in tight binaries, even though this trend is clearly confirmed only for separations of less than 40\,au, with~a disc frequency of only $\sim$30--40\% as compared to $\sim$80\% for single stars (Figure \ref{Krausdisc}).  Observational surveys also show that, even when present, disks are on average less massive in tight binaries \citep{harris12}.

This can have two major consequences on the planet-formation process. The~first one is that a truncated disk has a faster viscous evolution and should be more short-lived, leaving less time for planet formation \citep{mull12}. This could be especially troublesome for giant gaseous planets that need to accrete vast amount of gas before the primordial disk's dispersal, and~numerical studies have shown for example that the giant planet in the $\gamma$ Cephei binary (of separation $\sim$20\,au) would not have enough time to form in situ by the canonical core-accretion scenario. There is observational backing to this shorter lifetime of disks-in-binaries, with~\cite{kraus12} showing that the
the typical lifetime of disks in $\leq$40 au binaries is less than 1Myr,
as compared to the 5--10 Myrs typical of single stars.
The other potential consequence of disk truncation is that there might not be enough mass left to form jovian planets in close-in binaries. Studies focusing on the specific case of $\gamma$\,Cephei found that a truncated proto-planetary disk might contain just enough mass to form the observed planet \citep{jang08,mull12}, but~recent re-evaluations of this planet's mass to 9\,M$_{Jup}$ might challenge these optimistic results \citep{bene18}.

\subsubsection{Planetesimal~Accretion}

\begin{figure}[H]
\centering
\includegraphics[width=0.9\columnwidth]{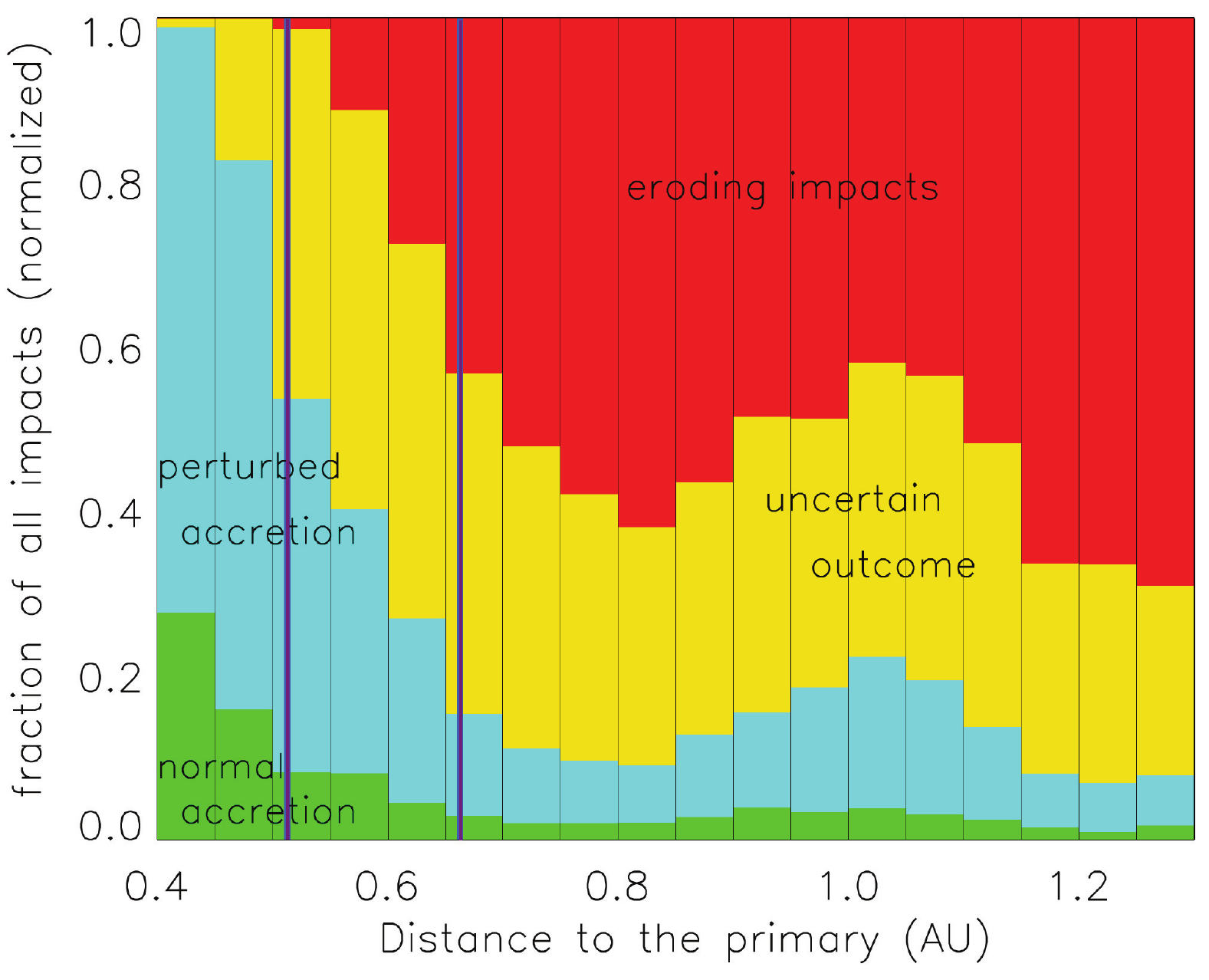}
\caption[]{The accretion behavior in a planetesimal disk around $\alpha$ Cen B, estimated with numerical simulations taking into account gas drag. The~relative importance of different types of collisional outcomes is displayed as a function of radial distance. \emph{Red}: impacts for which the impact velocity $\Delta v_{s_1,s_2}\geq v_{ero-M}$, where $v_{ero-M}$ is the threshold velocity beyond which an impact between two objects of sizes $s_1$ and $s_2$ always results in net mass loss. \emph{Yellow}: Uncertain outcome. The~erosion vs. accretion net balance depends on the physical composition of the planetesimals.
\emph{Green}: $\Delta v_{s_1,s_2}\leq v_{esc}$, where $v_{esc}$ is the escape velocity of the ($s_1,s_2$) pair. Accretion can here proceed unimpeded, in~a ``runaway~growth'' way, as~around a single star. \emph{Light blue}: $v_{esc}\leq\Delta v_{s_1,s_2}\leq v_{ero-m}$. Collisions result in net accretion, but~$\Delta v$ are high enough to switch off the fast-runaway growth mode.
The two thick blue lines denote the location of the inner limit of the ``optimistic'' and ``conservative'' habitable zone around the star. The~planetesimal size distribution is assumed to be a Maxwellian centered on 5\,km (modified from~\citep{theb09}).
}
\label{acentheb}
\end{figure}

The planet-formation stage that has been, by~far, the~most studied in the context of binaries is the intermediate stage leading from kilometer-sized planetesimals to Lunar-sized planetary embryos. This is probably because planetesimal accretion, controlled by their mutual gravity, requires very low encounter velocities, of~the order of a few m s$^{-1}$ for km-sized bodies, in~order to successfully proceed  (e.g., \citep{liss93}), and~is thus very sensitive to dynamical perturbations such as those coming from a stellar companion.
Various studies performed by the authors of this review showed that a crucial mechanism is in this case the coupling between the gravitational pull of the companion and the drag from the still-present primordial gas disk on the planetesimals. This coupling induces a strong size-dependent phasing of planetesimal orbits which reduces impact velocities among same-sized objects \citep{mascho00,paard08} but strongly increases those between non-equally sized bodies, leading to collision velocities far above the threshold allowing mutual accretion \citep{theb04,theb06}. For~any realistic distribution of planetesimal sizes, the~latter effect dominates the former, leading to a globally adverse 
consequence of the companion star on the accretion process. Assuming a small inclination between the binary and the circumstellar planetesimal disks tend to mitigate this effect, but~at the costly price of a drastic slowing down of the growth rate \citep{xie09,xie10a}. All in all, for~close-in binaries with separation $\sim$20\,au, models show that planetesimal accretion is prevented beyond $\sim$0.5--1\,au from each star \citep{theb08,theb09}. This makes the in situ formation of the HD196885b, $\gamma$ Cephei b or HD41004Ab planets problematic \citep{paard08,theb11}, and~also raises issues as to the potential formation of planets in the habitable zones within the $\alpha$ Centauri system (Figure \ref{acentheb}).

Recent analytical investigations showed that, if~the gas disk is massive ($M_d \geq 0.01 M_{\odot} $) and nearly axisymmetric ($e_d \leq 0.01$), then the disk's self-gravity could reduce encounter velocities among planetesimals to values low enough to allow their mutual accretion \citep{rafi13,rafi15,sils15}. However, numerical studies of gas disk dynamics in binaries showed that $e_d$ is probably never low enough (with~$e_d \geq 0.03$) for this mechanism to be effective \citep{kley07,marz09,marz12,zsom11,mull12}. Moreover, the~only preliminary  numerical study that has, to~this day, investigated planetesimal dynamics taking into account gas disk self gravity has found that disk gravity tends to increase impact velocities instead of reducing them \citep{frag11}. More detailed investigations are required to quantitatively assess this~issue.

\subsubsection{Late~Stages}

\begin{figure}[H]
\centering
\includegraphics[width=.7\columnwidth]{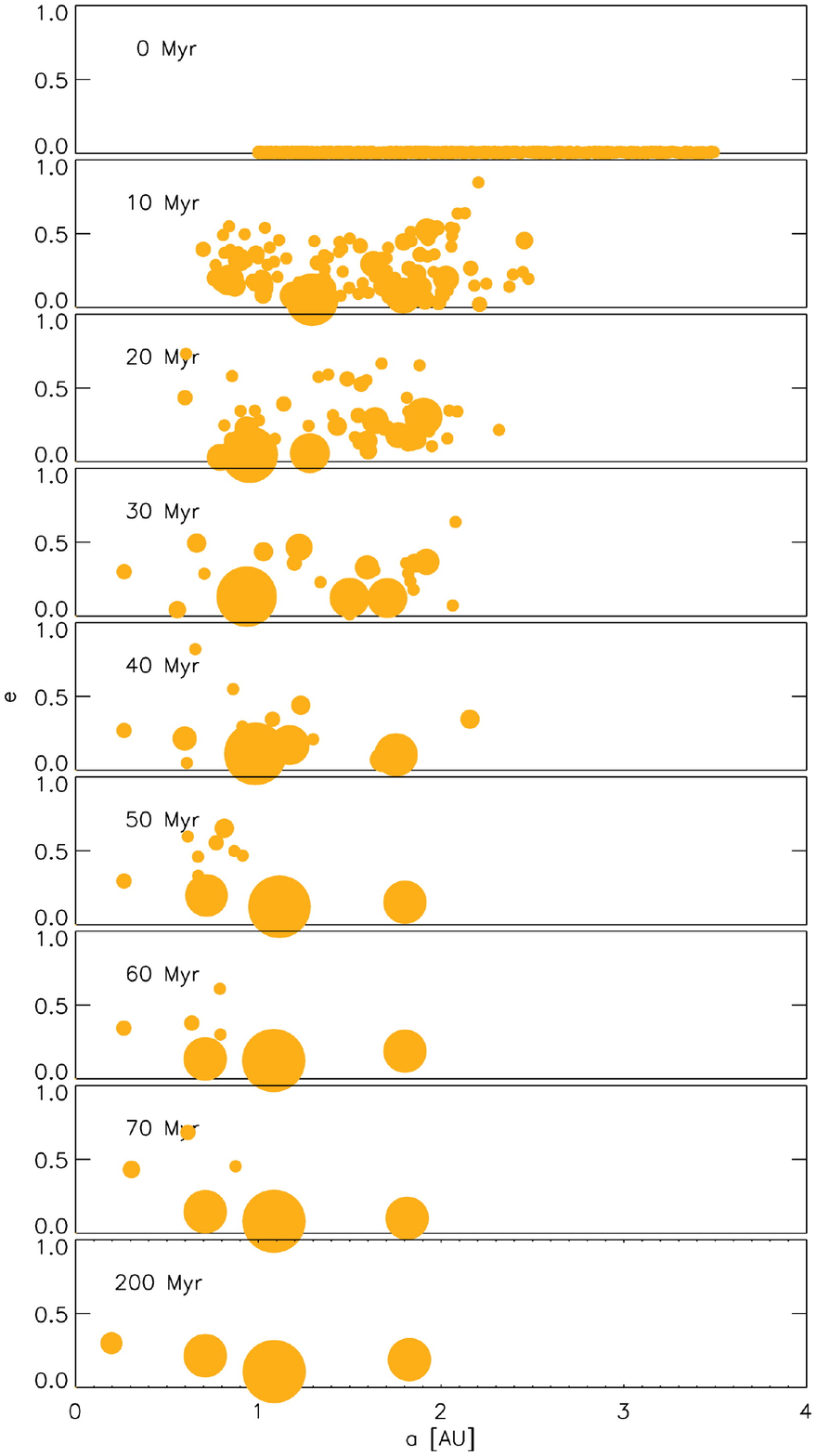}
\caption[]{Numerical simulation of the mutual accretion within an initial disk of lunar-sized embryos around $\alpha$ Centauri B. At~the end of the simulation, four terrestrial planets were formed. (Taken~from~\citep{guedes08}, courtesy of the Astrophysical Journal) }
\label{guedesim}
\end{figure}

The prospects for the last stages of planet formation, leading from lunar-sized embryos to fully formed planets, look much more promising. This mainly comes down to the fact that it is much more difficult to hinder the accretion of $\sim$1000\,km objects than that of kilometer-sized ones. The~companion star perturbations would indeed have to induce impact velocities exceeding $\sim$1 km s$^{-1}$ instead of a few m s$^{-1}$. This was confirmed by several numerical studies (see Figure~\ref{guedesim}) which showed that the mutual growth of large embryos is possible in mostly all the regions of orbital stability around each star \citep{barb02,quin02,quin07,theb04,haghi07,guedes08}. 
The one caveat is that, of~course, this last stage can proceed only \emph{if} the preceding stage of planetesimal accretion has been successful in forming lunar-sized~embryos.

\subsubsection{Alternative Scenarios?}

Given the potential concerns regarding some stages of the canonical planet-formation scenario, several authors  considered alternative explanations for the paradoxical presence of exoplanets in close-in binaries. One possibility is that these planets formed closer to their host star, in~regions more protected from the adverse influence of the stellar companion, and~later migrated outward. However, it seems that this migration is not large enough, limited to a radial displacement of $\sim$30\%, to~explain the HD196885 or $\gamma$ Cephei planets \citep{payne09}. In~the same family of scenarios, there is the possibility that these planets were injected in their current orbit after dynamical scattering with another planet \citep{marzari2005}. However, this would require the presence of another, more massive but yet undetected planet closer to the central star, which might be difficult to explain given that this second planet should in principle be easier to detect than the first one. On~a related note, reference~\cite{gong18} showed that such planet-planet scattering followed by tidal capture might transfer a planet from a circumbinary orbit (P-type) to a circumstellar one (S-type), effectively creating a tight planet-hosting binary. However, this process might only work for binary separations of less than $\sim$3\,au.

Another possibility is that planet formation follows a different channel in a binary environment.~References \cite{xie10b,paard10} for example showed that planetesimal growth might still be possible, but~in a different form, in~a high impact velocities environment. The~basic idea is that, while mutual planetesimal collisions are destructive and produce small  fragments, these produced fragments might be easily re-accreted by other planetesimals that were not destroyed by mutual impacts. These ``lucky seeds'' will then continue to sweep up such small dust-sized debris until they reach sizes big enough to be protected against destructive impacts with other planetesimals. However, these two studies remain very qualitative and this sweeping growth mode should be quantitatively investigated by more detailed models.
~Reference \cite{duch10} considers a more radical alternative, which is that most planets in $\leq$100\,au binaries are not formed by ``classical'' core accretion but by the concurrent gravitational instability scenario. The~main argument is that binary-truncated proto-planetary disks are more compact and thus more gravitationally unstable, and~also that companion perturbations could act as an additional trigger to start these instabilities. This explanation would also explain why planets in tight binaries tend to be larger than around single stars. However, this scenario does not seem to be backed by numerical models, which show that, on~the contrary, the~development of instabilities is hindered by the presence of a close-in companion \citep{nelson00,mayer10},  so that the presence of larger planets in tight binaries is probably not due to this alternative formation channel.

\begin{figure}[H]
\centering
\includegraphics[width=.9\columnwidth]{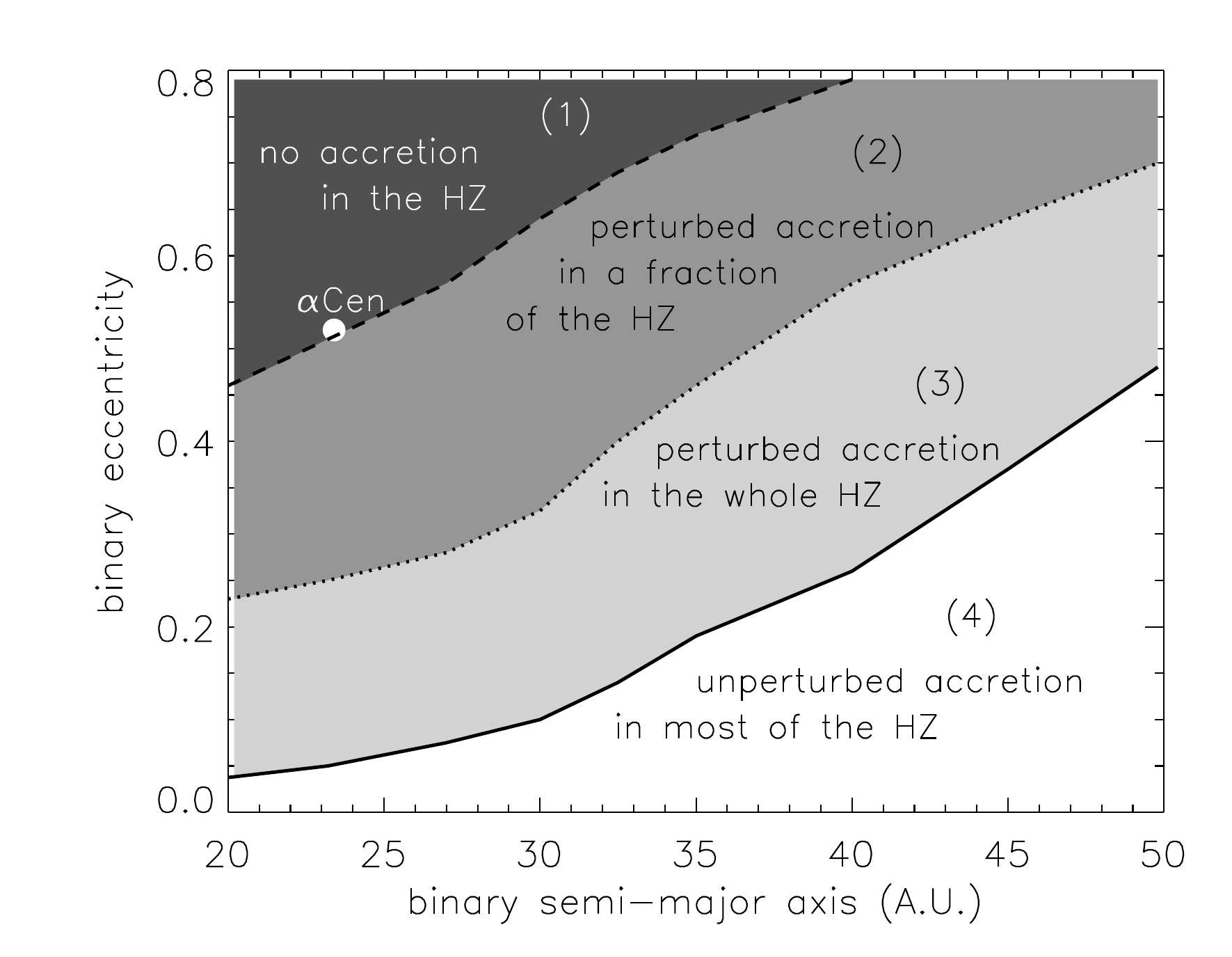}
\caption[]{Accretion vs. Erosion behavior of a population of kilometer-sized planetesimals in the habitable zone around $\alpha$ Centauri B when varying the binary's separation and eccentricity. The~white circle marks the current orbit of the binary. This shows that planetesimal accretion is possible if the binary has suffered a stellar encounter that shrank its initial wider orbit to its current value (taken~from~\citep{theb09}, courtesy of the MNRAS).}
\label{acenwide}
\end{figure}

The last family of solutions to the planets-in-tight-binaries paradox invokes stellar interactions that could have modified the dynamical environment around the host star \emph{after} the formation of the planet.~Reference \cite{marz07} showed that some tight planet-hosting binaries can be formed from unstable hierarchical triple systems, after~the ejection of the third star shrinks the central binary, which was thus initially wider and less disruptive for planet formation. Similarly,~reference \cite{malm07} showed that early stellar encounters in stellar clusters also lead, on~average, to~a shrinking of binary separations, so that the accretion-hostile binary configuration could come about {\it after} the formation of the planet (Figure \ref{acenwide}). More recently,~reference \cite{fragio19} estimated that there is a limited but non-zero possibility for a planet-hosting star to be injected into a pre-existing binary after an encounter with it, ejecting the former companion star and replacing it in the binary, at~a separation that is, on~average, smaller than that of the initial~binary.

\unskip

\section{Planets in P--Type Orbits: Circumbinary~Planets}

The number of known exoplanets on P--type orbits (circumbinary) is
smaller compared to those in S-type orbits, leaving their 
dynamical and physical properties less constrained. Most of them 
(at~present 10) were discovered by Kepler with the transit 
method while the more massive ones ($2.3 M_J < m < 10 M_J$) were 
found either by eclipsing binary timing
\citep{schwarz2011} around evolved stars or 
direct imaging. 
The most interesting among transiting circumbinary
planets is Kepler--47, the~only multi--planet system 
that has 3 exoplanets with masses ranging from 2 to about 43 
Earth masses moving in almost circular orbits, coplanar 
to the binary orbital plane, stable at least over 100 Myr and
with the inner and less massive planet right next to 
the dynamical stability limit 
\citep{orosz2019}. 
Almost all Kepler circumbinary planets found so far orbit their stars very 
close to the plane of
the binary in a prograde direction.  This coplanarity seems to be 
suggested also by the statistical analysis of~\cite{armstrong2014} showing that if planetary orbital inclinations 
were randomly distributed with respect to the binary orbital plane, 
then the inferred frequency of planets in 
circumbinary orbits should be exceptionally high compared to 
that around single stars. On~the other hand, if~coplanarity is 
the preferred orbital architecture, 
then the rate of occurrence of
circumbinary planets appears to be consistent with that of single~stars.

The three main features that seem to pop up from the current known population 
of circumbinary transiting planets are:

\begin{itemize}[leftmargin=10mm,labelsep=5.5mm]

\item[(1)]
	 a lack of planets around tight binaries ($T_{bin} < 7$ days) 
\citep{martin2018}

\item[(2)]
 a pile--up close to the inner dynamical stability~limit

\item[(3)]
 masses smaller than that of~Jupiter

\end{itemize}

The last two items face the exception of Kepler-1647 \citep{kostov2016}, which is
both the most massive (\mbox{$\sim$1.5 $M_J$}) P-type planet and the one with the widest orbit, with~a semi-major axis of 27 au well beyond the inner stability limit of the~system. 

The reason no P-type planet has been found around tight binaries
is possibly due to the formation path of such tight binaries, which 
might be  the outcome 
of the evolution of a triple system via the Kozai--Lidov mechanism
followed by dynamical instability that ejected the third star 
(see~\cite{moe2018} for recent modeling). As~for the pile-up of planets 
smaller than Jupiter, it is possibly due to the formation 
mechanism and subsequent evolution of planets in the circumbinary 
disk.

\subsection{Formation and Evolution of Circumbinary~Planets}

Several theoretical works have shown that the binary perturbations 
may create a hostile environment for forming a planet
close to the stellar pair.
~Reference \cite{meschia2012a} modeled the planetesimal accretion phase in
Kepler 16, the~first circumbinary planet detected with Kepler,
with an N-body code including the aerodynamic drag 
induced by an axisymmetric gaseous disk. He found that, although~
the size-dependent pericenter phasing of the orbits 
might allow planetesimal accumulation in a narrow region in between  1 
and 1.8 au, the~most likely region for the growth of planetary embryos 
lies beyond 4 au.
~Reference \cite{paard2012} improved this calculation by including, in~their model,
dust accretion onto planetesimals and extending the computations 
to the cases 
of Kepler-34 and Kepler-35. Their simulations confirm that 
planetesimal accumulation close to the central stars is strongly 
inhibited by the binary perturbations. 
An additional refinement to these N-body models was done 
by~\cite{lines2014}, which~included the effect of the
planetesimal disk self-gravity and considered an improved collisional
model. Their results confirm that, in~the case of Kepler-34, 
the inner region of the circumbinary disk is a hostile environment
for planetesimal accretion even for planetesimals as large as 
120 km. According to~\cite{rafikov2013}, the~region where planetesimal 
accumulation is forbidden may be reduced if we consider 
the effects of the gravity field of the gaseous disk 
on the secular evolution of the planetesimals. The~precession 
rates of the pericenters of the planetesimals are indeed faster 
once perturbed by a massive circumbinary disk
and a strong damping of the planetesimal eccentricity  is 
expected beyond 2--3 au from the stars, shifting inwards
the location of the accretion friendly zone, even if not 
enough to encompass the region where the planets are currently~located. 

\begin{figure}[H]
\centering
  \includegraphics[scale=0.45]{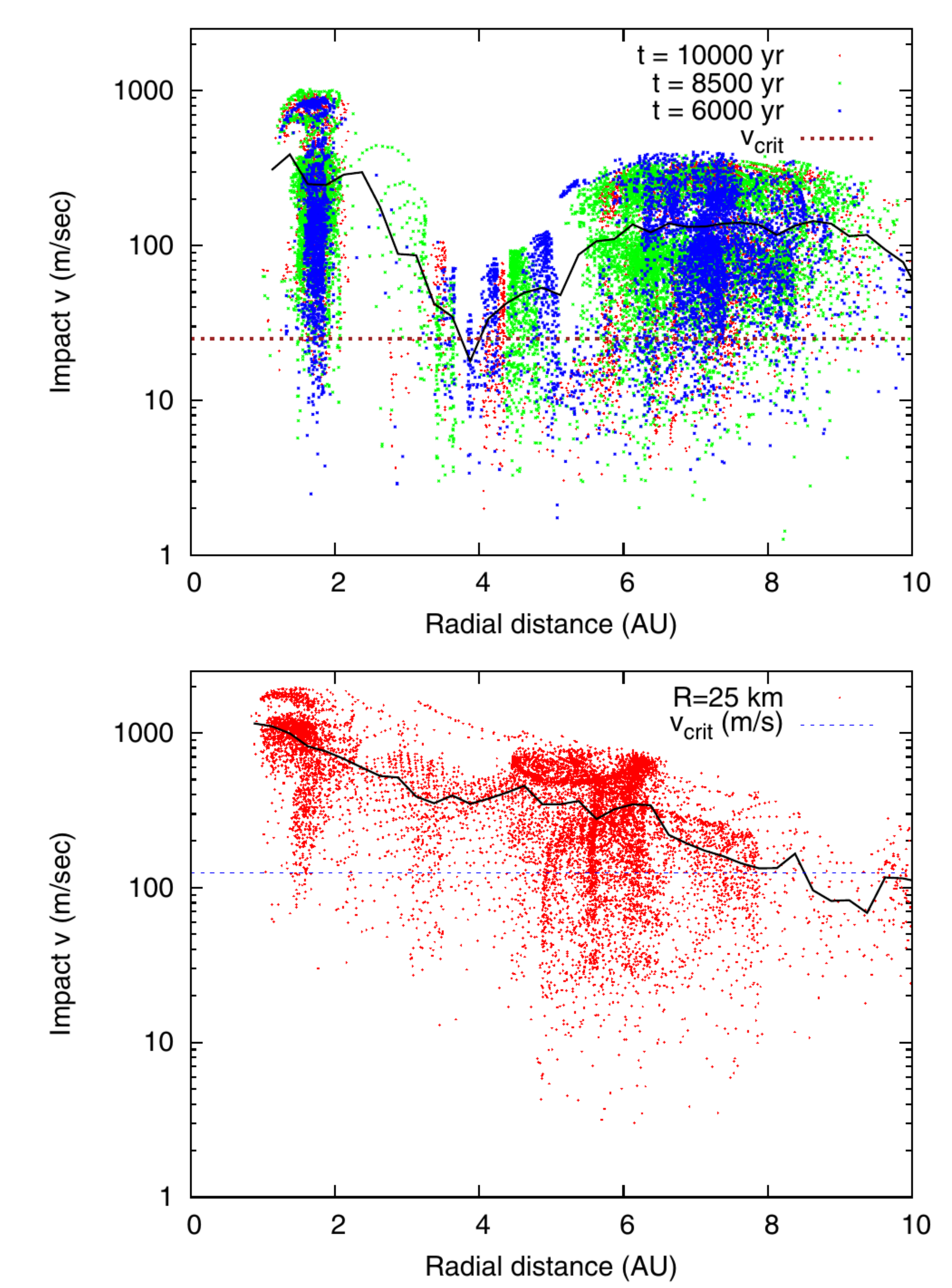}
	\caption{Figure from~\cite{marza_circum_2013} showing 
	at different evolutionary times  the
	relative impact velocities 
	for a putative planetesimal population
	in Kepler-16. 
	In the upper panel the impact velocities between equal size
	R = 5 km planetesimals, computed at different radial distances 
	from the baricenter of the binary, are compared  
	with the critical erosion velocity (black dashed line).
	The continuous black line shows the average impact 
	velocity for the population. In~the lower panel the 
	impact speeds are computed for larger planetesimals 
	R = 25 km in size. 
	}
 \label{circum_marza}
\end{figure}

In all the previous models the effects of gas drag on planetesimals
were computed assuming a fixed axisymmetric gas disk.~References  \cite{marza_circum_2008,marza_circum_2013}
relaxed this assumption and used an hybrid algorithm solving at the same time the evolution 
of the gaseous disk with the FARGO model \citep{masset2000} and
the N-body equations for the planetesimals including gas drag 
and the gravitational pull of the disk. They find that density waves 
and a significant disk eccentricity strongly affect the 
planetesimal orbital evolution, both~via a non-axisymmetric 
gas drag force and direct gravitational pull, leading to a 
higher dephasing of the pericenters and to higher mutual encounter velocities causing
erosive impacts. Figure~\ref{circum_marza} shows that,
for Kepler-16, the~whole region within
10 au from the central binary is very hostile to planetesimal
accretion due to the gravity from the eccentric gas disk. 
All these studies rule out the formation of a planet by core-accretion
close to its present location (near the binary inner stability 
limit), strongly suggesting a scenario in which the planet forms
further out in the disk and subsequently moves
with traditional migration processes to its current~orbit.

\unskip

\subsection{Migration Towards the Inner~Hole}

The pile-up of most circumbinary planets just outside the inner stability limit despite the difficulty 
of forming them in these highly perturbed regions suggests that they formed further out 
and later migrated inward by interaction with the circumbinary disk. 
Indeed, if~a massive Jupiter-like planet forms in a circumbinary disk, 
it carves a gap and it drifts inward in the regime of Type II migration. 
At some stage of its inward drift, the~planet-carved gap will reach the outer edge
of the large cavity formed by tidal pull around the central binary, merging with it  to
form a single wide gap in which both the 
stellar pair and the planet are embedded \citep{nelson2003}. The~planet 
will then continue to migrate until it reaches the 4:1 MMR with the 
binary's orbit and, at~that point, either it stalls close to the MMR 
or it may follow different unstable paths \citep{nelson2003}.
According to~\cite{sutherland2016},
in most cases the planet is ejected out of the system on 
a hyperbolic orbit, filling
the population of free floating planets, or~it may impact either the 
secondary star (more frequently) or the primary contributing to develop 
a difference in star metallicity between the two
due to their different amount of planet
pollution. More complex is the evolution in the case of less massive planets not opening a gap 
in the disk. Type I migration drives them close to the stellar pair, 
but when they get close to the inner border of the disk hole, 
their final location is determined by a balance between the gravitational 
pull of the stars and the interaction with the disk in which they 
are still embedded. According to the modeling of~\cite{kleyhaghi2014,kleyhaghi2015} 
the main relevant parameter is then the thermal profile of the 
disk, which may significantly influence the last stages of evolution 
of these lower-mass planets as they approach the 4:1 or 5:1 MMRs. In~addition, 
they find that, to~prevent the scattering of the planet out of 
the system, a~low density disk is required and, possibly, the~present of
a second planet in the~system. 

\section{Dynamics and Stability in S-Type~Orbits}

The main players determining the dynamical evolution of 
planets orbiting in the S-type configuration are:

\begin{itemize}

\item The interaction with 
the circumstellar disk.  In~the early phases of evolution of the system it
may significantly change the architecture of 
the planetary system \citep{kleynelson2010}.
\item Mean Motion Resonances (hereinafter MMR) between the planets
and between the planets and the binary companion. 
\item Secular perturbations and secular resonances, linear and non-linear 
	\citep{mich2004,libert2005}, 
with the main frequencies of the system. 
\item The Kozai dynamics when significant mutual inclinations are present
	among the bodies \citep{naoz2016}. 
\item Planet--planet or planet--star scattering if the system becomes 
	temporarily unstable \citep{marzari2005}. 
\end{itemize}

All these mechanism may act, together or at different times, 
to determine the evolution of a system of planets in binaries. 
In particular, to~determine the outer limit of stability of a 
planet orbit~\citep{holman99,marzari_gallina2016}, secular and mean 
motion resonances are the most active players. Superposition 
of resonances may determine large areas of instability where 
planets cannot survive for a long enough time~\citep{mudrik_wu2006,deck2013}. 
Secular perturbations, by~increasing the average eccentricity of 
the planet(s) orbit, may drive them into these chaotic region 
caused by resonance~overlap.

\subsection{Secular~Evolution}

A simple model for the secular perturbations of the binary companion 
on the orbit of  a circumstellar planet is given by~\cite{heppenheimer1978}
which derived an averaged disturbing function in the orbital elements 
of the planet and binary star which leads to the solution for the 
planet $h,k$ non-singular orbital elements of the form:
\begin{equation}
	\frac {dh}  {dt} = g_s (k - e_F), \qquad \qquad 
	\frac {dk}  {dt} = -g_s h
\label{eq:heppen1}
\end{equation}

\noindent
where $g_s$, the~frequency of circulation of the pericenter longitude of the planet, is given by:
\begin{equation}
	g_s= \frac {3} {4} \mu \frac {a_p^3}{a_b^3} 
	\frac {n_p}{(1 - e_b^2)^{3/2}}
\label{eq:heppen2}
\end{equation}

\noindent
The forced eccentricity $e_F$ is computed as:
\begin{equation}
        e_F= \frac {5} {4} \frac {a_p}{a_b} 
	\frac {e_b}{(1 - e_b^2)}
\label{eq:heppen3}
\end{equation}

\noindent
with $a_p, n_p$ the semi-major axis and mean motion
of the planet, $a_b, e_b$ the semi-major axis and eccentricity of 
the binary companion, $m_0$ the mass of the central star, $m_b$ 
that of the secondary and 
$\mu = \frac {m_0}{m_b}$ their ratio. From~Equation~(\ref{eq:heppen3})
it follows that $e_F$ is proportional to $a_p$ and it increases 
while approaching the outer 
stability limit of the planet while the frequency $g_s$ also depends on 
the  mass ratio $\mu$. 

\begin{figure}[H]
\centering
  \includegraphics[width=7 cm]{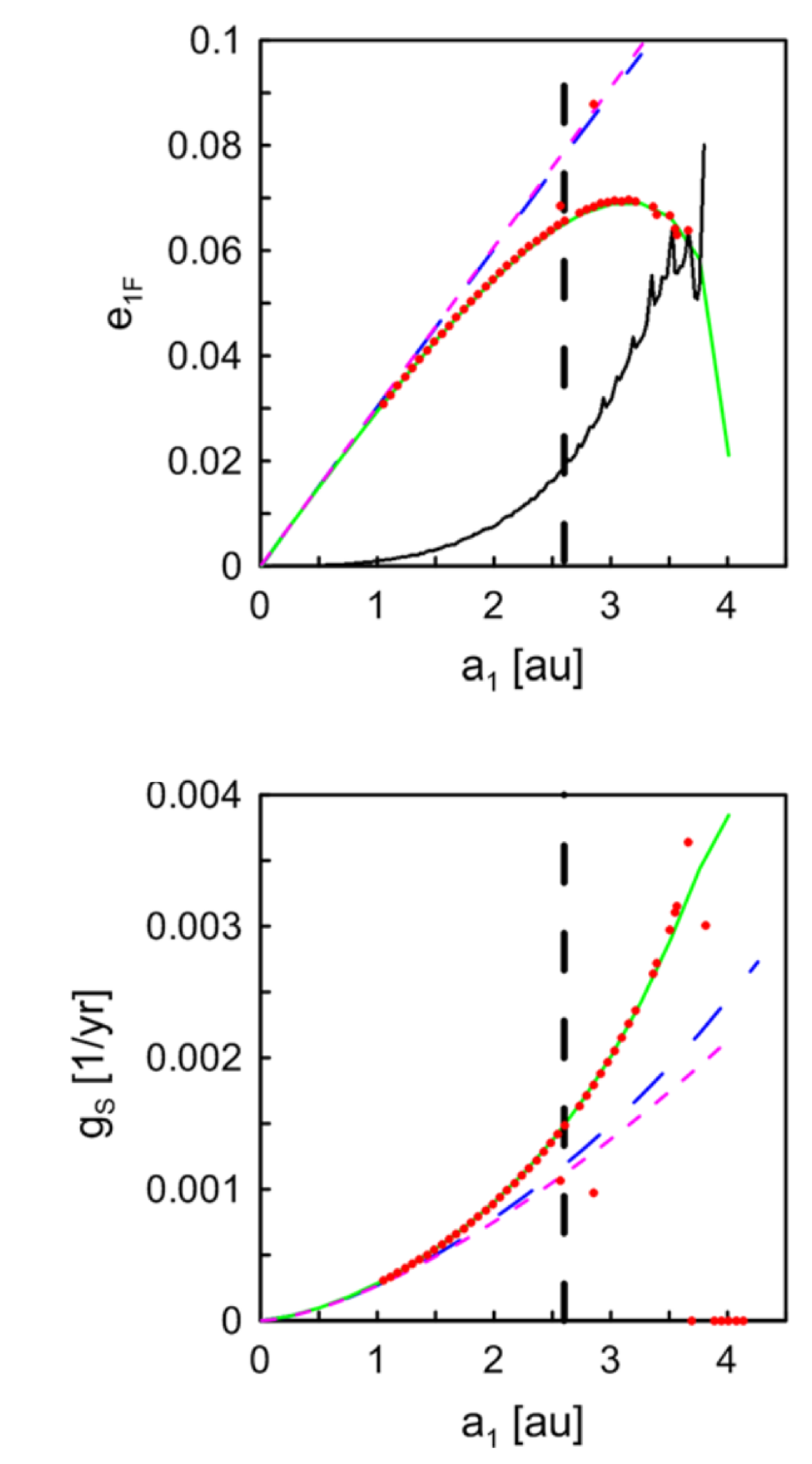}
  \caption{\label{andra}
Figure taken from~\cite{andrade2016}. The~values of 
	$e_F$ (top panel) and $g_s$ (bottom panel) 
	are computed for the planet in 
	HD 196885 (the vertical dashed line shows its location).
	The red dots are reference outcomes of numerical 
	simulations~\cite{holman99}, the~magenta curve is derived from~\cite{heppenheimer1978}, the~blue curve from the 
	first order model of~\cite{andrade2016} while 
	the green curve is the second order model of~\cite{andrade2016}.
	The black curve shows the amplitude of the short period 
	variations of the eccentricity.}
\end{figure}

Higher order theories are needed when either the ratio $a_p/a_b$ or $e_p$,
or both,
are not small. Reference~\cite{andrade2016} has shown  that, under~these circumstances, 
second order theories are more accurate in computing both $e_F$ and 
$g_s$ close to the outer border of stability. 
As illustrated in Figure~\ref{andra}, the~model of~\cite{heppenheimer1978}
in general predicts too high values of $e_F$ and slower oscillation 
frequencies $g_s$
respect to the outcome of a direct numerical integration while 
the second order analytical model follows more closely the 
numerical~solution. 

\unskip

\subsection{Stability Limit for One~Planet} 

A planet in a binary system is affected both by secular and 
resonant perturbations which may interact and lead to 
instability beyond a given value of $a_p/a_b$. 
The threshold value of $a_p$, for~a given value of $a_b$, is 
called the critical 
semi-major axis $a_c$ and it  
has been estimated for 
different values of $a_p/a_b$, $e_b$ and mass ratio 
$\mu$ by
~\cite{rabl88,holman99} via a large 
number of numerical simulations. The~more accurate
semi-empirical 
formula for $a_c$,  given by
~\cite{holman99}, is:
\begin{align}
a_c =\, & [ ( 0.464 \pm 0.006) + (-0.380 \pm 0.010) \overline{\mu}   \\  \nonumber
      & + (-0.631 \pm 0.034) e_b + (0.586 \pm 0.061) \overline{\mu} e_b  \\  \nonumber
      & + (0.150 \pm 0.041) e_b^2 + (-0.198 \pm 0.074) \overline{\mu} e_b^2 ] a_b,
\label{eq:HW}
\end{align}

\noindent
where the mass ratio is defined as $\overline{\mu} = m_b / (m_0 + m_b) $. 
The source of instability beyond $a_c$ is related to the growth of 
the forced eccentricity $e_F$
due to the secular perturbations which becomes higher for larger values of 
$a_p/a_b$ (Equation (\ref{eq:heppen3})). This drives the planet into the 
region where mean motion resonances (MMRs) overlap leading to 
chaos \citep{mudrik_wu2006}. The~critical semi-major axis estimated by~\cite{holman99} is, however, a~lower estimate. A~more accurate numerical 
approach based on the FMA (Frequency Map Analisys,~referneces \cite{lask93, sine96, marz03}) by~\cite{marzari_gallina2016} has shown~that: 
\begin{itemize}
\item 
the value of $a_c$ computed by~\cite{holman99} 
underestimates the real stability limit
and stable planetary orbits can be found beyond 
$a_c$. 
\item some planets may be trapped in low order resonance (3:1, 5:2)
with the companion star, far beyond $a_c$, and~be stable over 
a long timescale. 
\item for large values of $e_b$, large unstable regions can be 
found within $a_c$ due to non-linear secular resonances with the 
binary companion.
\end{itemize}

\begin{figure}[H]
\centering
  \includegraphics[width=9 cm]{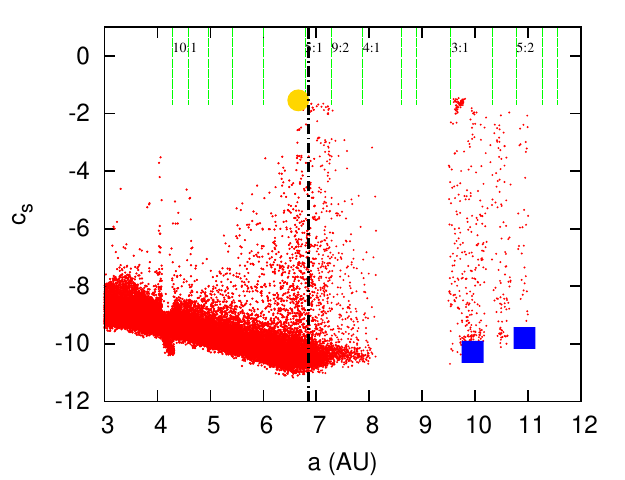} 
  \includegraphics[width=9 cm]{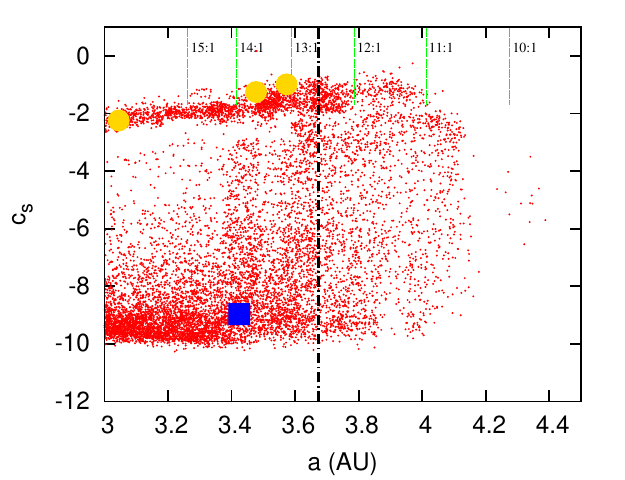}
  \caption{\label{galli}
Figure taken from~\cite{marzari_gallina2016}.  
The FMA analysis is performed on a Jupiter-size single planet 
orbiting in a
binary system with $a_B = 25$ AU and $e_B = 0$ (top panel) and 
$e_B = 0.4$ (bottom panel). The~diffusion
index $c_s$ of the main secular frequency of the planet 
is drawn vs. its semi-major axis.
Small values of $c_s$ means low diffusion, while large values
connote chaotic orbits. The~black dash-dotted line marks
the critical semi-major axis computed from the empirical
formula of~\cite{holman99}. The~green dashed lines show the
location of mean motion resonances between the planet and the
companion star. The~blue squares show stable cases whose
orbits were numerically integrated over 4 Gyr, the~yellow
circles unstable ones. 
        }
\end{figure}

In  Figure~\ref{galli} this behavior is illustrated for a circular 
and eccentric ($e_b=0.4$)  binary with a semi-major axis $a_b= 25$ au. 
The FMA diffusion index $c_s$ is plotted for different values of 
the planet semi-major axis. Low values mean stability, high values 
suggest chaotic behavior. In~both cases the stability limit
exceeds that derived from~\cite{holman99} while unstable 
cases can be found within it, in~particular in the eccentric 
case. In~addition, in~the circular case, some putative planets
are trapped in stable mean motion resonances with the companion 
star well beyond the stability~limit.

\subsection{Stability of Multiple Planet~Systems}

When two or more planets are present in the system, the~situation is 
more complex since, in~addition to the resonances with the binary, 
mean motion and secular resonances develop among the planets that
may overlap with those due to the companion star. 
As a consequence, we do not expect that a system of two 
planets in binaries to be necessarily stable if both planets 
orbit within $a_c$. Reference \cite{pilat2016}
studied the location of secular resonances and chaos in a system
with two planets, focusing their investigation on the system HD 41004. 
They consider different binary configurations and test the stability 
of a putative inner planet perturbed also by the giant planet
detected in the system. The~stability of the inner planet is 
studied with the maximum eccentricity criterion and a 
semi-analytical method is outlined to determine the location of 
secular resonances within $a_c$ with a reduced computational load.~Reference \cite{marzari_gallina2016} has instead applied the FMA method to 
test the influence of a binary companion on the Hill stability 
limit for two giant planets orbiting within $a_c$. They 
found that the minimum separation between the two planets for stability
is still marked by the 3:2 mean motion resonance between them
but that the level of chaoticity close to this threshold is 
higher in a binary system. In~addition, 
they derived a semi-empirical 
equation giving the minimum semi-major axis $a_b'$  of a binary orbit 
compatible with a stable two-planet system orbiting the main star. 
This computation is  
different from the calculation of the critical $a_c$ 
for a single planet~\cite{holman99} since it requires that 
two planets are present in the system. The~value of 
$a_b'$ is given by:
\begin{equation}
a_b' = \frac {10.} {1 - e_b} (1. - 0.54 (1 - \mu^2)) \cdot a_1
\label{eq:MG}
\end{equation}

\noindent
where $\mu = m_b/m_0$ and $a_1$ is the semi-major axis of the 
inner planet. This equation has a limited range of validity
i.e., $ e_b < 0.6$, $10 < a_b < 100$au, $0.1 < \mu <1$ and
$2 < a_1 < 4$au but it can be exploited to predict the presence 
of an additional planet when one has been already detected. 
It is an approximated formula based on a large number of numerical
tests but it does not have an accuracy estimate due to the 
complexity of the numerical~interpretation. 

\subsection{Kozai Evolution in Presence of Inclined~Binaries}

Planets in binaries may have their orbital plane tilted with respect to 
that of the binary. This may occur either because the rotation 
axes of the stars in the binary are misaligned at the moment of 
their formation \citep{hale94,Chen2008,ratzka2009,jensen2014}
or because a dynamical mechanism like planet-planet scattering
\citep{marzari2005} has lead to large inclinations
of the planets with respect to their original orbital plane (possibly the equator
of the central star). In~either cases, the~misalignment has important 
dynamical consequences on the long term evolution of the planetary
orbits related to the Kozai mechanism. 
The classical Kozai evolution is usually described by the quadrupole 
approximation of the Hamiltonian that, neglecting the Keplerian
terms, is:
\begin{equation}
	F_q = -\frac {e^2} {2} + cos^2(i) + \frac {3} {2} e^2 cos^2(i) +
	\frac {5} {2} e^2 ( 1-cos^2(i)) cos (2 \omega )
\label{eq:FQ}
\end{equation}

\noindent
where $a,e,i,\omega$ are the orbital elements of the planet in 
a reference frame attached to the binary orbit. Since 
$F_q$ does not depend on the node $\Omega$, the~component of 
the angular momentum along $z$
is constant and this leads to the classical conservation law:
\begin{equation}
	J_z = \sqrt{(1 - e^2)} cos(i)
\label{eq:Jz}
\end{equation}

\noindent
If the initial mutual inclination between the planet and the 
binary companion is higher than approximately 39$^\circ$ 
the oscillations in eccentricity, always in anti-phase with those 
in inclination, may become wide even if the initial eccentricity
of the orbit was small. This behavior was invoked to explain the 
high eccentricity of the giant planet orbiting the star 16 Cygn B
\citep{holcygn1997}. If~the planet formed in a circular orbit
but on a plane misaligned by more than 39$^\circ$ with respect to that of the binary, 
its orbit would switch between low-eccentricity and 
high-eccentricity orbits. During~the high eccentricity phase, a~strong
tidal interaction with the central star may cause an inward 
migration of the planet, the~so-called 'Kozai migration' \citep{wu2003}.

The classical formulation of the Kozai mechanism 
assumes that the planet is handled as a
massless test particle. If~the planet is massive, than~the expression
for $F_q$ changes but the main properties of the evolution are 
similar. However, if~the coefficient:
\begin{equation}
	\epsilon = \frac {(a/a_b) e_b}  {1 - e_b^2}
\label{eq:naoz1}
\end{equation}

\noindent
is not negligible, in~particular when the eccentricity of the 
binary is high, then the quadrupole approximation is not accurate 
and the octupole term must be considered \citep{naoz2016,litnaoz2011}.
Interesting behaviors appear when the octupole term is considered, 
in particular the planet (either as a test particle or as a 
massive body) may undergo flips from prograde to retrograde orbits 
and the large oscillations in both eccentricity and inclinations
may become chaotic. The~large eccentricity achieved during the 
oscillations may also lead the planet to impact the star, unless~
some mechanisms like the General Relativity periastron precession
and tides suppress the eccentricity~excitations.

\section{Dynamics and Stability in P-Type~Orbits}

Circumbinary orbits are perturbed by the 
tidal gravity field of the stellar couple which 
establishes a web of MMRs close to them.
This web is responsible for:

\begin{itemize}
\item the unstable region surrounding the two stars due 
to resonance overlap,
\item halting the planet migration in the early phases of 
the evolution of the system by trapping the planet in a
stable resonance. 
\end{itemize}

\subsection{A Secular~Theory}

 Reference \cite{morinaka2004} developed a secular theory for bodies 
orbiting around a stellar binary which leads to equations similar
to those derived for S-type orbits. The~non-singular variables
$h,k$ are solution of the following differential equations:
\begin{equation}
	\frac {dh}  {dt} = A k -B cos(\varpi_b), \qquad \qquad 
        \frac {dk}  {dt} = - A h -B sin(\varpi_b)
\label{eq:heppen1}
\end{equation}

\noindent
where
\begin{align}
	A & = \frac {3} {4} \frac {n_b^2} {n} \left (\frac {a_b} {a} 
	\right )^5 \overline{\mu} (1-\overline{\mu}) \left(1 + \frac {3} {2} e_b^2 \right ) \\
	B & = \frac {15} {16} \frac {n_b^2} {n} \left (\frac {a_b} {a}
	\right )^6 \overline{\mu} (1-\overline{\mu}) (1 - 2 \overline{\mu}) \left(1 + \frac {3} {4} e_b^2 
	\right )e_b
\label{eq:morinaka1}
\end{align}

\noindent
and $\overline{\mu}
= m_b /(m_0 + m_b)$. The~forced eccentricity is given by:
\begin{equation}
	e_F  = \frac {B} {A} = \frac {5} {4} (1 - 2 \nu) \frac {a_b} {a}
	\left( \frac {1 + \frac {3} {4} e_b^2} 
	{1 + \frac {3} {2} e_b^2} \right )
\label{eq:morinaka2}
\end{equation}

\noindent
The frequency $g_s$ of oscillation of the
free eccentricity is equal to $A$. Contrary to the 
case of S-type orbits, 
both $e_F$ and $g_s$  increase while approaching the baricenter of 
the stars since both quantities are inversely proportional 
to the semi-major axis 
of the body. The~increase of $e_F$ for smaller radial distances 
facilitates the entry into the chaotic region surrounding the 
two stars and due to resonance~overlap. 

In the initial stages of a circumbinary system, when the gaseous 
disk is still present, the~secular evolution of the system has to 
include the gravity of the disk. As~shown in~\cite{rafikov2013}, 
the term $A$ in Equation~(\ref{eq:morinaka1}) becomes $A + \dot{\varpi_d}$
where the additional term $\dot{\varpi_d}$ is given by:
\begin{equation}
	\dot { \varpi_d} \sim - \frac {1} {4} n \frac {M_d} {M_b} 
	(\frac {r} {r_0})^{1/2}
\label{eq:rafi1}
\end{equation}

\noindent
assuming an initial superficial density profile for the disk:
\begin{equation}
	\Sigma (r) \sim \frac {M_d} {4 \pi r_0^2} ( \frac {r} {r_0})^{3/2}
\label{eq:rafi2}
\end{equation}

\noindent
with $M_d$ mass of the disk, $M_b$ the binary mass equal to the 
sum of the masses of the two stars and $r_0$ a reference radial 
distance. In~addition, the~
term $B cos(\varpi_b)$ is no longer constant because 
$\varpi_b$ changes with time. 
The precession of the binary orbit forced by the disk potential is
given by:
\begin{equation}
	\dot {\varpi_b} \sim - \frac {1} {8} n_b \frac {M_d} {M_b} (
	\frac {a_b^3} {r_0^{1/2} r_{in}^{5/2}} )
\label{eq:rafi3}
\end{equation}

\noindent
where $r_{in}$ is approximately equal to $2 a_b$, i.e.,~twice the 
binary semi-major axis. 
The forced eccentricity in this scenario assumes a value given by:
\begin{equation}
	e_F = \frac {B} {A +  \dot {\varpi_d} - \dot {\varpi_b}}
\label{eq:rafi4}
\end{equation}

\noindent
and its value depends on both $\dot \varpi_d$ and 
$\dot \varpi_b$ 
while the precession frequency becomes $g_s = A + \dot {\varpi_d}$, 
faster than in the case without the disk potential. While this 
correction to the secular theory is expected to be important for 
the planetesimal evolution, it is probably less relevant when 
dealing with fully formed planets since,  at~that evolutionary
stage, the~disk is significantly less~massive. 

\subsection{The Inner Unstable~Hole}

Bodies orbiting a stellar pair are influenced by its tidal time-periodic
gravitational potential.~ References \cite{rabl88,holman99,morinaka2002} explored 
the effects of these potential on putative planets orbiting 
the binary proving the existence of an unstable chaotic region 
surrounding the two stars. Planets ending up in this region have their eccentricity pumped up and they have a close encounter with the binary.
This gap surrounding the stars has been shown to form also for 
circumbinary disk by~\cite{artilub94}.
Following the 
same procedure adopted to outline the stability region for 
S-type orbits,~\cite{holman99} calculated a critical semi-major
axis beyond which orbits are dynamically stable. It can 
be computed as:
\begin{align}
a_c = \, & [ ( 1.60 \pm 0.04) + (5.10 \pm 0.05) e_b + (-2.22 \pm 0.11) e_B^2   \\  \nonumber
      & + (4.12 \pm 0.09) \overline{\mu} + (-4.27 \pm 0.017) e_B \overline{\mu}  \\  \nonumber
	&  + (-5.09 \pm 0.11) \overline{\mu}^2 + (4.61 \pm 0.36)  e_B^2 \overline{\mu}^2 ] a_b, 
\label{eq:HW2}
\end{align}

\noindent
As stated above, the~mechanism responsible for the dynamical 
instability within $a_c$ is the resonance overlap of MMRs with 
the binary while the forced eccentricity 
$e_F$ (Equation \ref{eq:morinaka2}) favors the entry into the 
chaotic region. If~multiple planets are present in
the system, the~behavior is also influenced by 
mutual secular perturbations, in~addition to those with the 
binary, and~mutual MMRs. In~this scenario General Relativity is
not expected to be relevant since the planets orbit 
always far enough from the star. Beyond~the critical value 
of $a_c$ there may be islands of instability, in~particular 
for high value of $e_b$, due to mean motion resonances 
\citep{sutherland2016}.

Reference \cite{martin2016} has shown that 
the Kozai evolution is not important in this 
scenario, at~least for 
planets with masses smaller than 5 $M_J$ and a binary
mass ratio $\mu > 0.1$. The~stellar 
binary system is not significantly affected by the perturbations of 
the outer body in this~case. 

\section{Perspectives}

An important step towards a better understanding of the formation and 
evolution of planets in binaries is to enrich the sample of 
known planets, in~particular for the circumbinary ones. At~present,
only a dozen of them are known and it is still not clear if some of their properties,
like the pile-up at the inner edge of the stable region or the 
overall low mass, are real characteristics due to the perturbations of 
the binary or if they are due to under-sampling and bias. 
As for planets on S-type orbits, what is crucially needed is a large-scale
survey that is both unbiased against binaries and target stars that are close
enough for their potential multiplicity to be well established beforehand. Another
potentially interesting survey would be one specifically targeting a statistically significant
sample of all the nearest~binaries.

Another important issue that requires a significant observational effort
concerns the orientation of the binary orbit with respect to the 
equator of the main star and the potential misalignment of the 
planetary orbit. The~early study by~\cite{hale94} pointed toward a
transition from alignment to misalignment around 30--40 au of separation, but~updated observational
investigations, as~well as theoretical works, are now clearly needed.
Locating the transition 
from alignment to misalignment may help the interpretation of the 
dynamical features of planets in binaries and, in~particular, 
give an estimate of the importance of the Kozai mechanism in 
sculpting planetary systems in~binaries. 

From a theoretical point of view, additional work is needed on one important stage of the planet-formation scenario that has
so far barely been investigated in the context of binaries, that~is the coagulation of the first condensed grains into
to kilometer-sized planetesimals. The~pioneering work by 
\citep{nelson00} has for instance shown that the heat generated by spiral shock waves might inhibit dust coagulation
in $>50$ au binaries.  In~addition, 3-D simulations
by~\cite{picogna2013} showed that strong hydraulic jumps occur 
at the shock front of spiral waves which are expected to affect
the vertical distribution of dust particles and their accretion
process by different merging mechanisms (collisional~or instability driven 
accumulation, dust traps etc., see~\cite{johansen2014} for a review). 
The presence of spirals will not only affect the timescale and initial 
size distribution of planetesimals but also the following stages 
of the core-accretion process like the fast pebble accumulation of 
planetary cores  \citep{ormel2010} which is 
expected to accelerate the growth of giant~planets. 

Regarding the stage that has so far proven to be the most sensitive to binary perturbations, i.e.,
the mutual accretion of planetesimals into Lunar-sized embryos, a~crucial step forward would be the
development of a numerical model coupling the dynamics of planetesimals and of the gas disk, and~taking
into account both the gas drag and the effect of the disk's self gravity. This would allow
quantitatively assessing the respective importance of the concurring accretion-favoring and
accretion-inhibiting effects triggered by the response of the gas disk to binary~perturbations.

\vspace{6pt}

\authorcontributions{The authors have equally contributed to the 
writing of the paper.}

\funding{This research received no external funding}

\conflictsofinterest{The authors declare no conflict of~interest.} 

\reftitle{References}

\end{document}